# THE DESIGN AND FABRICATION OF PLATFORM DEVIEC FOR DNA AMPLIFICATION

*Chao-Heng Chien, Hui-Min Yu*

Mechanical Engineering Department, Tatung University
40 Chung Shan N. Rd. Sec. 3 Taipei, Taiwan

## ABSTRACT

Thermalcycler were extensively used machine for amplify DNA sample. One of the major problems in the working time was that it spent most of time for cooling and heating. In order to improve the efficient, this study presented a novel method for amplify DNA sample. For this concept, the DNA sample in the silicon chamber which was pushed by a beam through three temperature regions around a center and then the DNA segments could be amplified rapidly after 30 cycles. The polymerase chain reaction platform was composed of thin-film heaters, copper plates, DC powers, and temperature controllers. The photolithography and bulk etching technologies were utilized to construct the thin-film heater and DNA reaction chambers. Finally, 1μL 100bp DNA segment of *E. coli K12* was amplified successfully within 36 minutes on this PCR platform.

## 1. INTRODUCTION

Since the MEMS technologies had been developed, it was widely applied for many territories, such as optics, electric, biology and so on. In biotechnologies, microfluidic chip and microarray chip were the major research in bio-MEMS for recent years. It took many advantages, for example, portable, miniature, short reactive time and automation. These years, the MEMS technology was also employed to fabricate the chip for amplifying DNA, such as continuous-flow PCR [1,2] flow-through thermocyclers [3,4], or closed chamber PCR-chip [5,6].

In the past decade, PCR was a very important role in the field of molecular biology. It was a very powerful technique that could amplify a little amount of nucleic acid to generate plentiful material for diagnose. Typically, PCR included three steps which were denaturation of double-stranded DNA, annealing of oligonucleotide primer pairs, and extension of new DNA strands catalyzed by DNA polymerase at temperatures of approximately 95°C, 55°C and 72°C, respectively [7]. Following the cycles of temperature zones, theoretically, a DNA segment got amplified 2n times after n cycles of PCR [8]. Fig. 1 showed the reaction process of PCR. In fact, the enzyme efficiency was decayed after each cycle. Thus, the efficiency of DNA amplification was $A=(1+e)^n$ where e approximated 0.8-0.9 [4].

In general, PCR cycles were performed in a thermocycler that took 2~3 hours for 30 cycles. As a result of time was almost spent on cooling and heating during the reactive time. In order to expedite the PCR efficiency, a different method of temperature transition between three temperature zones and low reaction volume were presented in this research.

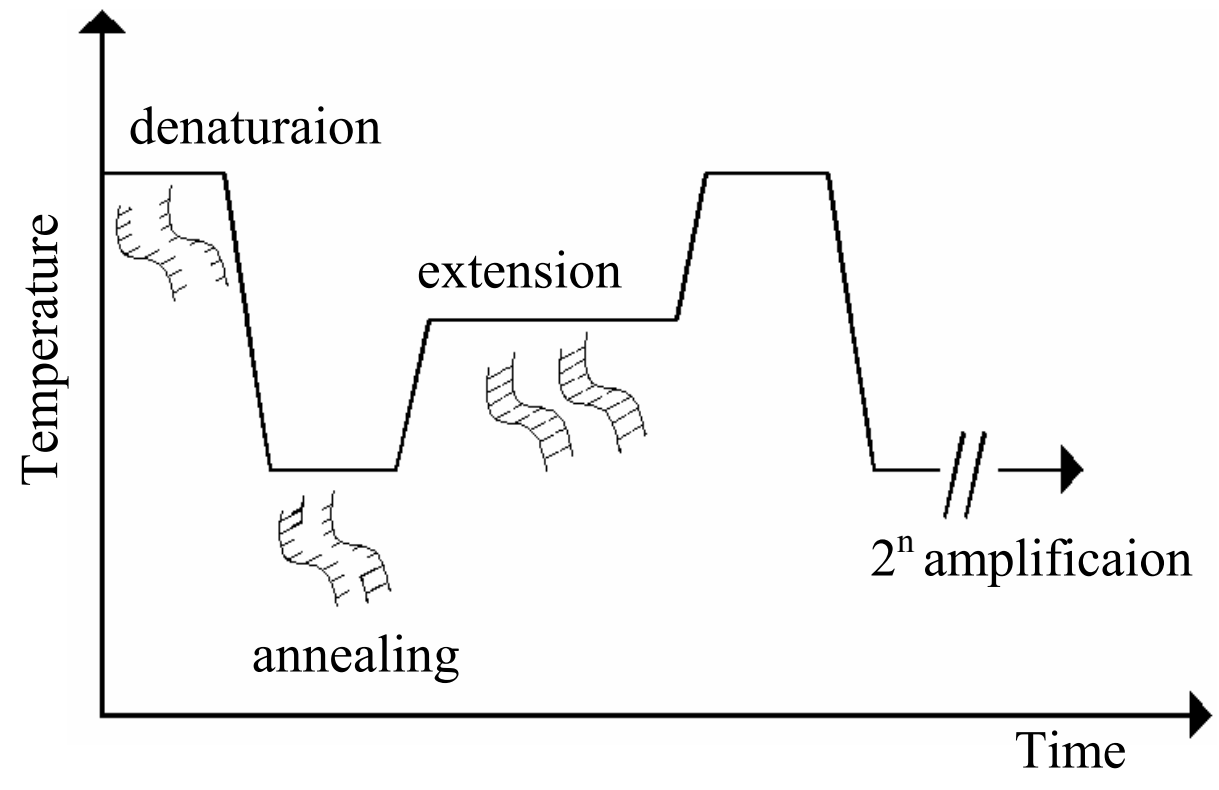

Fig. 1. The DNA amplification process of PCR [4].

## 2. DESIGN OF PCR PLATFORM

The PCR platform in this research consisted of two parts that were the rotator device and the heater device shown in Fig. 2. In the rotator device, both of the DC motor and counter were constructed on the PMMA frame. The rotation rate of tappet was controlled via the DC motor. The DNA chamber on the heater-based was driven by the tappet, and the cycle number was counted. The heater device consisted of four components which were the base, heat resistance plate, heater, and copper plate. The heat resistance plate was the isolation layer between the copper heater and the base to induce the most of heat flow of the heater transmitted into the copper plates. The copper plate with 2mm thickness above the thin-film heater was to make uniform temperature distribution. The copper was deposited and patterned on the glass wafer to be the thin-film heater.

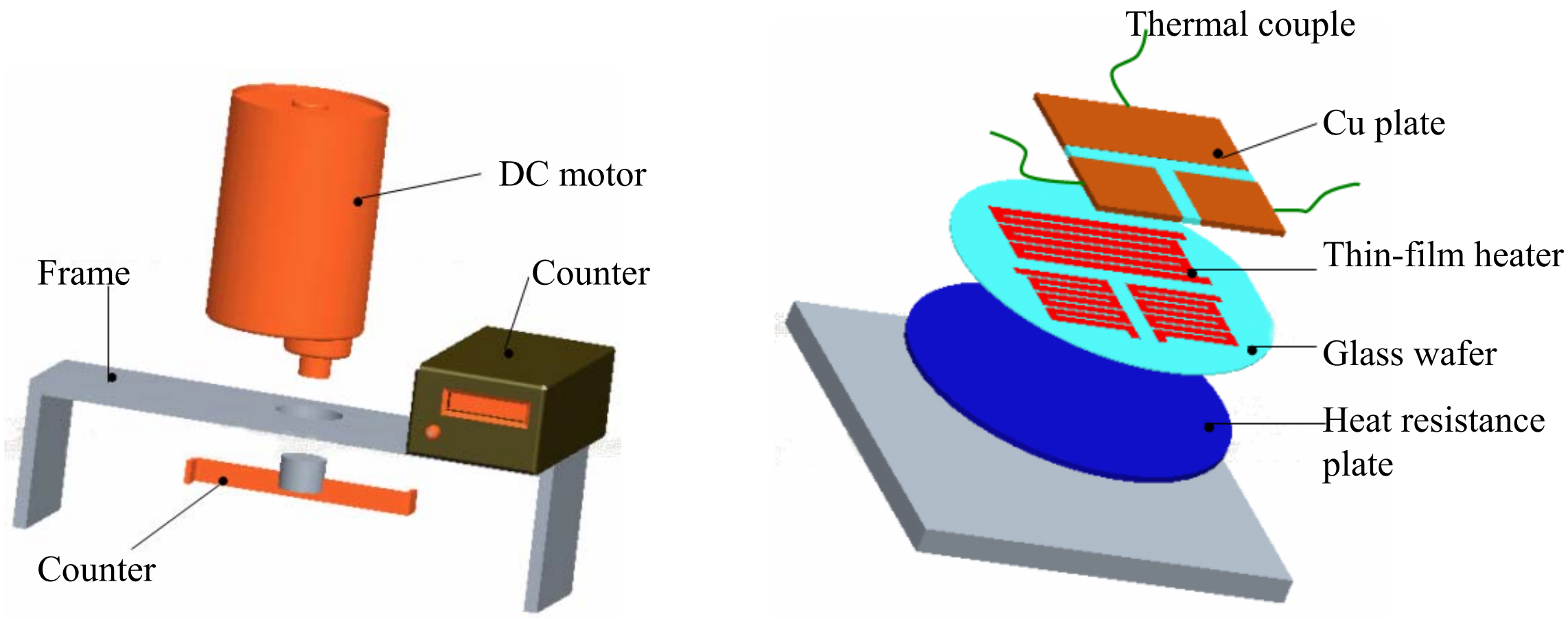

Fig. 2. The explode scheme of PCR platform. (a) The rotation device. (left) (b) The heater-based. (right)

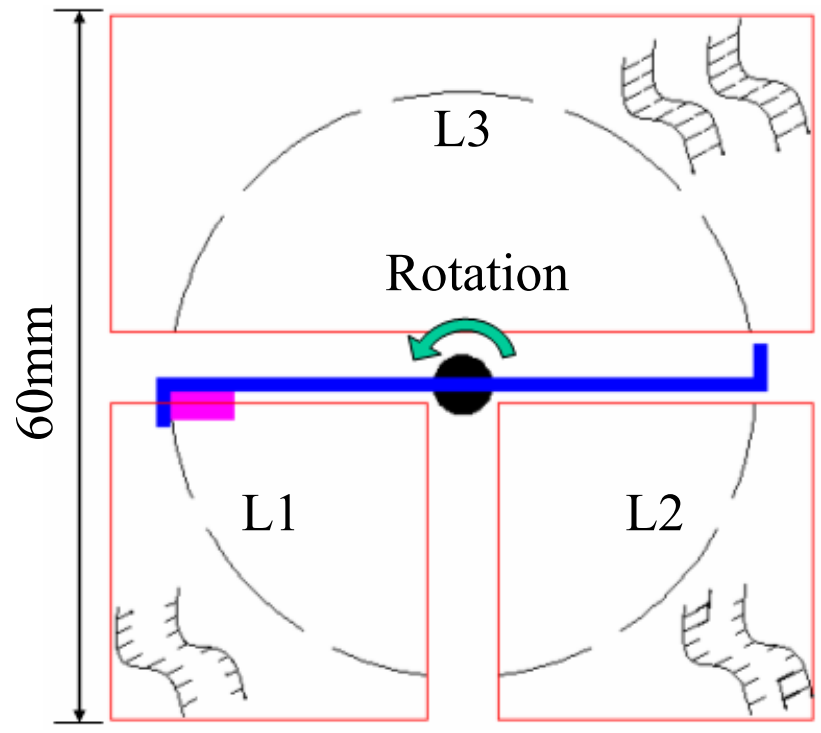

Fig. 3. The trace of DNA chamber on the heater-based with the ratio of trace L1： L2： L3 = 1： 1： 2. Unit：mm

The trace of DNA chamber was described in Fig. 3. The unclockwise rotation of DNA chamber went through the 95°C, 55C° and 72°C regions per cycle. The ratio of trace was L1： L2： L3 = 1： 1： 2. According to the different sample, the ratio of the trace varied with the location of the tappet center.

## 3. FABRICATION OF HEATERS AND TEMPERATURE CONTROL

In the temperature control system, the proportional integral derivative (PID) controller was used to regulate the temperature. The thermal couple was employed as the temperature sensor and inserted into the Cu plate shown in Fig. 2(b). The Cu of 2000Å thickness was deposited on the glass wafer to be as the thin-film heaters. Fig. 4 showed the fabrication process of the deposition. The Ti/Cu/Ti was deposited 500Å, 2000Å, and 500Å, sequentially. The purpose of titanium material deposited was to be the adhesive layer and to avoid the oxidation of the copper heater. At last, the pad was soldered with the wire to complete the thin-film heater.

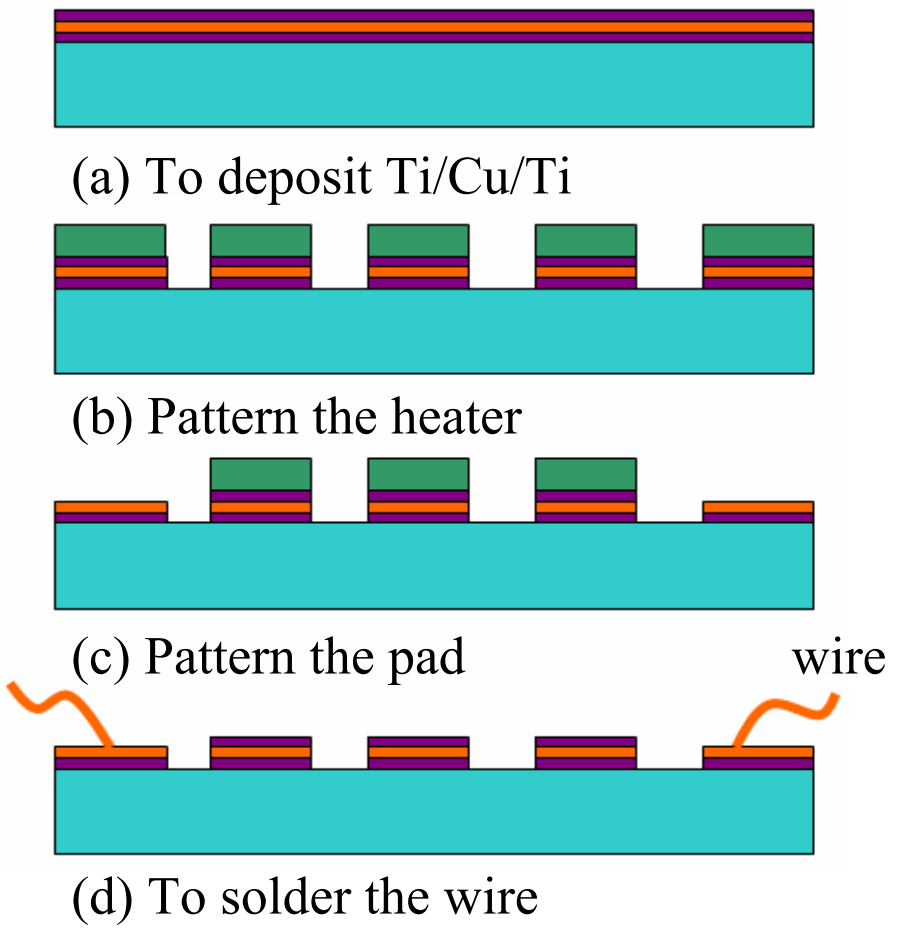

Fig. 4. The fabrication processes of the thin-film heater on glass wafer.

## 4. FABRICATION OF DNA CHAMBERS

The correct DNA sample was dependent of the uniform temperature distribution on the PCR sample. The accelerated heating/cooling rate decreased the time in the amplified DNA. Therefore, the silicon material was chosen to be fabricated as the DNA chamber because it had high thermal conductivity induced a better temperature distribution and saving reaction time. Fig. 5 showed the fabrication processes of DNA chamber. The <1 0 0> silicon wafer of 525μm thickness with the double side silicon nitride was the substrate. The photolithography technique was utilized to pattern chamber before removing the $Si_3N_4$ by dry etching (RIE) and showed in Fig. 5(b). The 3-D chamber was etched of 300μm depth by wet anisotropic KOH solution to make the side wall of 54.74° with <100> [9], then the upper $Si_3N_4$ was removed and shown in Fig. 5(c-d). Owing to the PCR would be inhibited in the silicon and silicon nitride surface of the DNA chamber [10]. Formation of SiO2 layer or chemical silanization of the chamber surface had been applied to increase compatibility in such silicon-based devices [5]. Therefore, the thermal oxide was form with 0.2μm thickness on the chamber surface in order to avoid the inhibition of the enzyme reaction on the silicon material, showed in Fig. 5(e). The complete DNA chamber was shown in Fig. 6. The volume of chamber designed was 4.5μl.

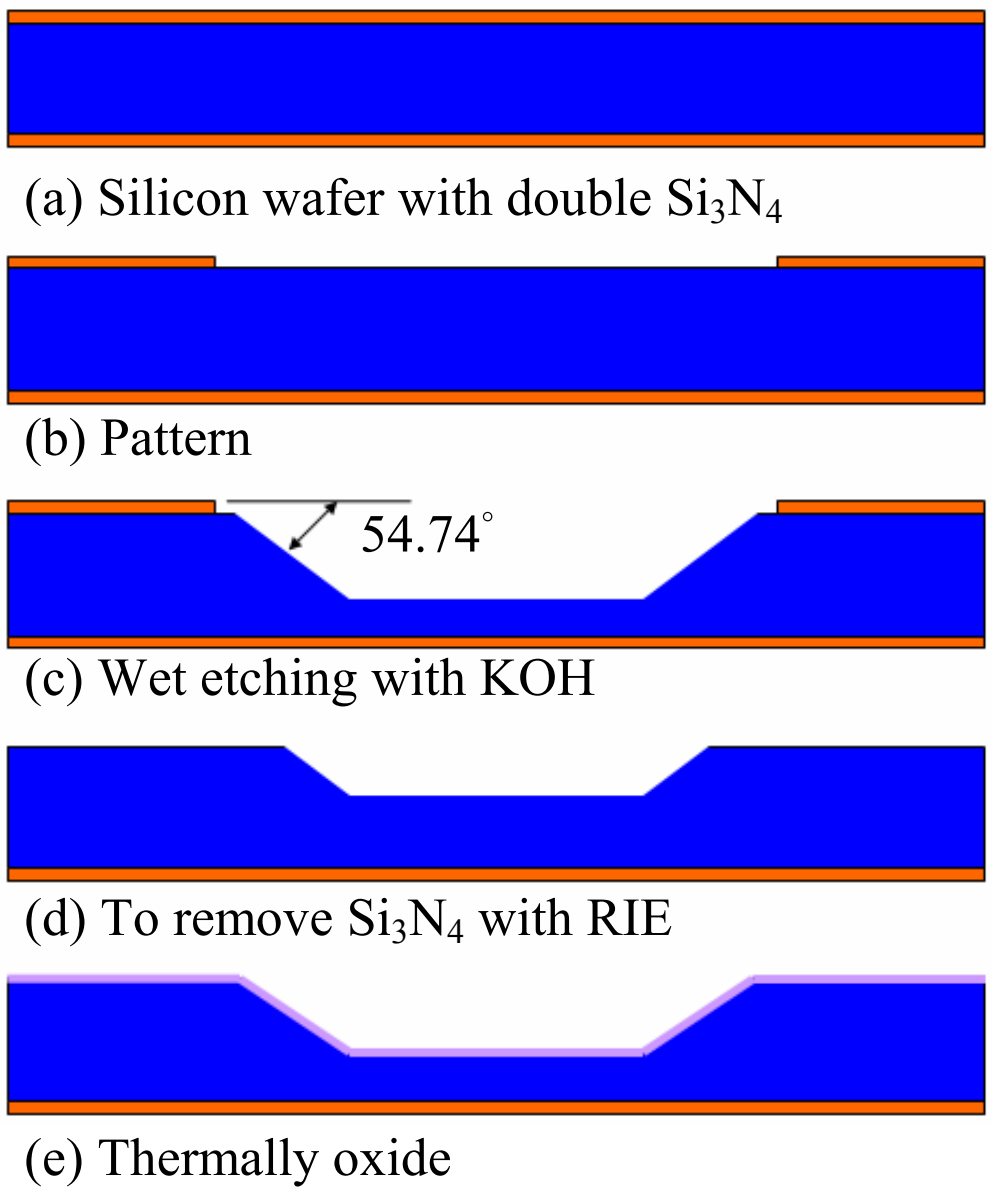

Fig. 5. The fabrication processes of DNA chamber

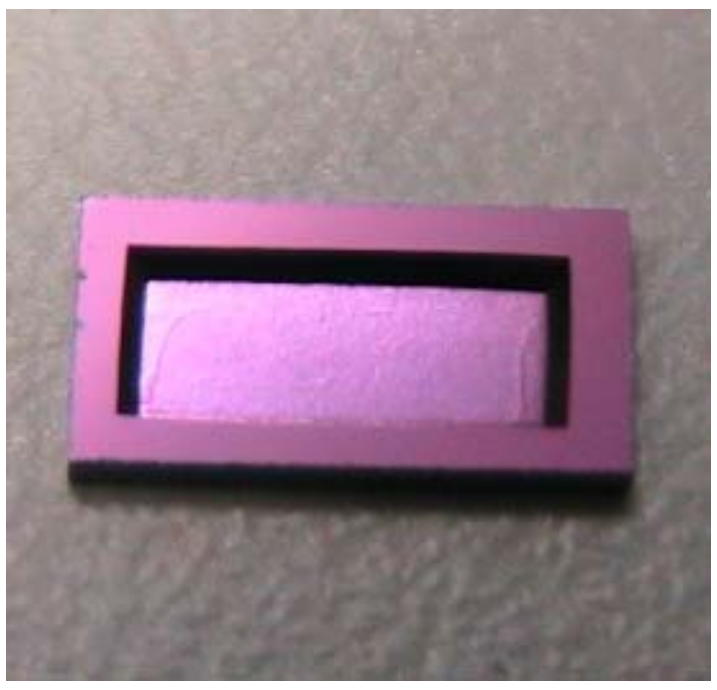

Fig. 6. The single DNA chamber with outer size was 5.8mmx2.8mm and inner size was 5.4mmx2.4mm.

## 5. EXPERIMENT AND RESULTS

In this session, the platform was to be tested for amplifying *E. coli K12* genomic DNA. The result was also compared with the commercial thermal cycler.

### 5.1. Materials

The PCR reaction mixture contained 1ng DNA template, 0.2mM dNTP, 0.48μM for each primer, and 1U TaqEnzyme (Super-Therm). The sequences of primer set to amplify a 100 bp target segment of 16s rRNA gene where 5'-GGA TTA GAT ACC CTG GTA GT-3' (forward) and 5'-CTT GCG RYC GTA CTC CCCA-5' (reverse). The *E. coli K12* genomic DNA used as the template was amplified.

### 5.2. Experiment

Fig. 7 showed the platform apparatus. In the experiment, the three DC power were supplied to the three thin-film heaters for 95°C, 55°C and 72°C, individually, and the 4.51W, 1.35W, and 3.74W of power consumption correspond to 95°C, 55°C and 72°C. The DNA chamber was cleaned and dried in the bake oven with 120°C for 5 minutes. The sample was dropped with 1μl into the DNA volume. Afterward, the mineral oil covered on the chamber to prevent evaporation of the sample when the boiling point of mineral oil was higher than the sample. The chamber was on the 95°C Cu plate to be denatured with 2 minutes at the first step in polymerase chain reaction. Then, the tappet was started to rotate the DNA chamber through the three temperature regions at 1 rpm. After 30 cycles were executed, the chamber should be kept in 72°C for 4 minutes in the final step. The total reaction time was 36 minutes including 15s denaturation at 95°C, 15s annealing at 55°C and 30s extension at 72°C for each cycles. The collected samples were analyzed by slab gel electrophoresis. Fig. 8 showed the results of gel electrophoresis of PCR productions. Lane1 was the marker, lane2 was the PCR platform, and lane3 was the commercial thermal cycler. As a result of lane2, the 100bp of PCR productions was observed and there were no specific products to be found. Thus, the PCR platform could amplify DNA segment rapidly. However, the amount of PCR product was less than product of the traditional PCR thermal cycler so the fluorescence intensity was not so good.

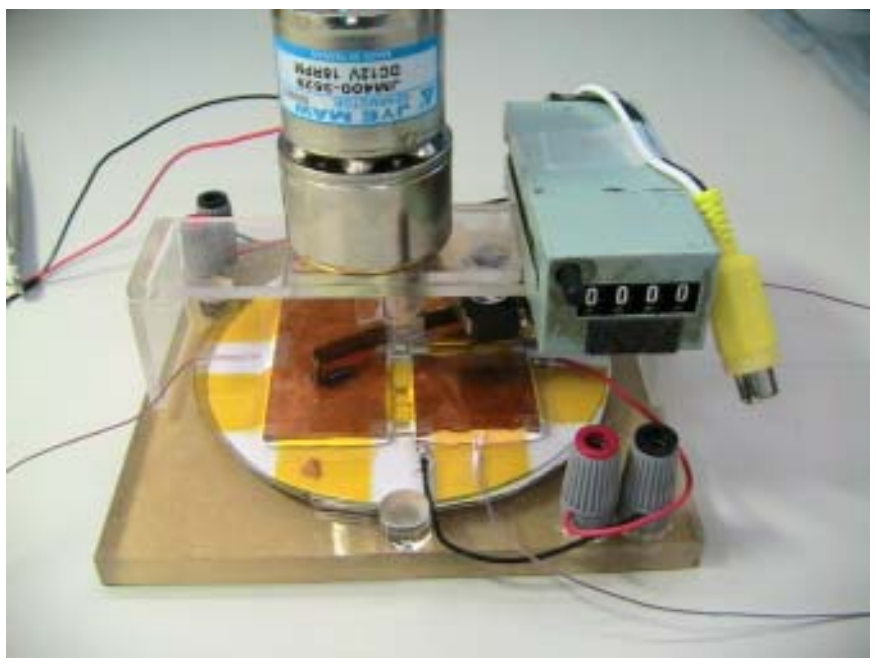

Fig. 7. The PCR platform apparatus.

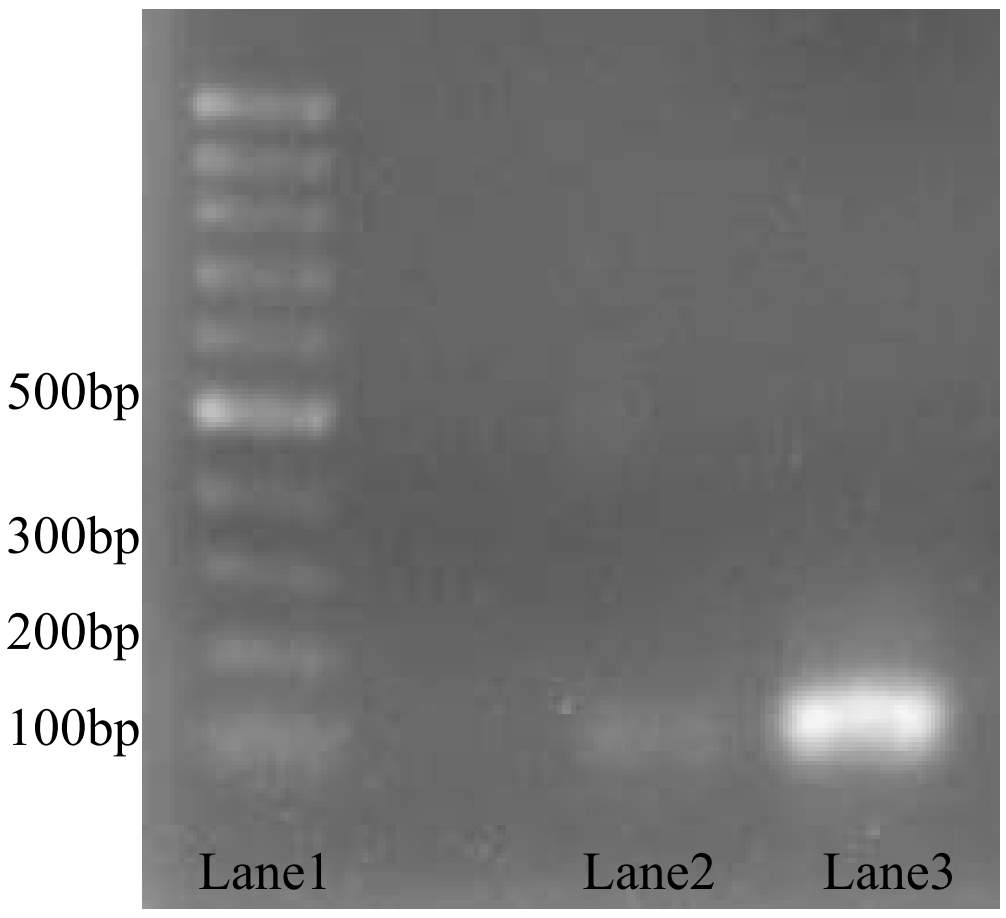

Fig. 8. The results of gel electrophoresis. Lane1: The marker. Lane2: The PCR platform (volume ~0.8μl). Lane3: The commercial thermal cycler.

## 6. CONCLUSION

A simple apparatus for rapid DNA amplification was developed in this research. The major components of the PCR platform included DC motor, tappet, heater plates, PMMA frame, and the temperature control system. The DNA chamber of 300 $\mu$m depth was fabricated using the wet anisotropic KOH solution to make the side wall of 54.74°. An excellent thermal transition rate of chamber was obtained due to the high thermal conductivity of silicon. Finally, a 100 bp segment of *E. coli K12* was demonstrated successfully for amplification. The total reaction time took 36minutes with 1μl volume for 30 cycles. However, the PCR platform was an easy and low-cost apparatus to be fabricated. It could be used in common Lab.